\documentclass{article}

\usepackage{arxiv}

\usepackage[utf8]{inputenc} 
\usepackage[T1]{fontenc}    
\usepackage{hyperref}       
\usepackage{url}            
\usepackage{amsfonts}       
\usepackage{graphicx}

\usepackage{subfig}
\usepackage{bm}
\usepackage{amsmath}
\usepackage{float}

\title{One-dimensional relativistic particle as a movable cellular automaton }
\author{ {A.\,M.~Pupasov-Maksimov} \\
	Department of Mathematics\\
	Universidade Federal de Juiz de Fora\\
	\texttt{pupasov.maksimov@ufjf.br} \\
	}

\begin{document}
\maketitle





\begin{abstract}
The one-dimensional dynamics of identical discrete elements that combine the properties of newtonian mechanical particles and cellular automata are investigated. It is shown that the motion of a cluster of combined discrete elements, which is the simplest observable object of the model, leads to the Feynman chessboard model whose continuous limit gives the Dirac equation. 
\end{abstract}

\maketitle

\section{Introduction}
Dirac's well-known recipe for creating a physical theory is to find the right application for a beautiful mathematical formalism \cite{dirac1940xi}:

{\it The method is to begin by choosing that branch of mathematics which one thinks will form the basis of the new theory. One should be influenced very much in this choice by considerations of mathematical beauty. It would probably be a good thing also to give a preference to those branches of mathematics that have an interesting group of transformations underlying them, since transformations play an important role in modern physical theory, both relativity and quantum theory seeming to show that transformations are of more fundamental importance than equations. Having decided on the branch of mathematics, one should proceed to develop it along suitable lines, at the same time looking for that way in which it appears to lend itself naturally to physical interpretation.
}

The method proposed by Dirac gradually began to dominate theoretical physics, and the complexity and abstractness of mathematical constructions rose to unattainable peaks.
At the same time, it is known that quite simply constructed dynamical systems can lead to complex evolution and structures. A striking example is the dynamics of cellular automata in the Conway game "Life" \cite{conway1970game}. The formulation of the rules is very simple and does not require a high level of abstraction, but the resulting structures and their interactions are much richer.
Thus one can accept the Dirac method and study systems generated by cellular automata as candidates to fundamental physical theories \footnote{Besides fundamental questions, cellular automaton are developing tools to solve practical problems \cite{wolf2004lattice}. }. Starting from some simple elements and interaction rules one can arrive to a world, populated by reach variety of structures with complicated behaviors \cite{wolfram1983statistical,wolfram2002new}. Following this line, Steven Wolfram started his physics project with very promising preliminary results \cite{wpp}. For instance, some computational universes demonstrate properties of relativistic space time. Quantum behavior also can appear due to dependence on the order of application of rules.  

Note that our experimental and computational tools (computers) are based on the physical reality. If computational nature of universe is considered, one can ask what is a platform to conduct necessary computations.  That is, which physical conditions may give rise computations to emerge spontaneously \cite{langton1990computation}.

 In the present article we discuss the continuous cellular automaton model presented on the website \textit{\bf http://insighttotal.org},  \cite{instotal}. The model pretend  to explain why exists such a fundamental constant as the upper limit for the speed at which conventional matter, energy or any signal carrying information can travel through space. It is also revealed the nature of dark energy, the equivalence of mass and energy, the  formation of observed matter and physical fields and quantum character of nature at the microscopic level.

Here we consider a simplified one dimensional model. There are impenetrable particles with finite size which we call Imps. 
Imps can be combined together and form a cluster. Such combinations should be interpreted as connections which lead to informational communications. 
Thus Imps are movable cellular automate \cite{psakhie1995method}, and basic rule is that connections between Imps have finite duration. In this case one can distinguish two configurations: free Imps which form a gas and clusters of Imps. Cluster of Imps during their life-time are subject to random walks between free Imps. Such random walks of a cluster can be connected to the Feynman chessboard model which leads to the one dimensional Dirac equation.

\section{Continuous cellular automaton} 
\subsection{Generalities} 
As a rule, when talking about cellular automata, some discrete systems with a discrete time step are assumed.  However, the concept of movable cellular automata is often used to describe the motion of complex mechanical systems \cite{psakhie1995method}. Movable cellular automata (MCA) obey mechanical laws in the form of differential equations. Each element has also some internal state which can be changed as well as relationships of pairs of automata can be switched.
Such combination allows to use computational advantages of both discrete elements methods and cellular automata methods. Various physical phenomena  (mass mixing, penetration effects, chemical reactions, intensive deformation, phase transformations,  accumulation of damages, fragmentation and fracture) are modeled within this framework.

Below we will study a one dimensional universe populated by simplest movable cellular automata. We call a single cellular automaton by {\it Imp}, as a contracted version of Impenetrable. For computational purposes we fix an auxiliary Galilean coordinate system and each Imp 
is characterized by its position and velocity \footnote{Note that this coordinate space is an auxiliary object to simplify formulation of the model}. Imps have non-zero size they collisions should be processed by cellular automata rules, which we describe below. Separated Imps form a gaseous state, while connected Imps form clusters.

However, gas of Imps differs from an ideal gas, a simplest model of statistical physics. Note that introducing finite volume of particles and some interactions one will arrive at the problem of a real gas which is almost intractable by analytical methods. 
We choose a simple interaction between colliding Imps which can be considered as resonances manifested by temporary capture of colliding particles \footnote{However, it is better to interpret such collisions as temporal connections of simple computational modules}.
The resulting cluster has the same momentum as the total momentum of colliding particles, while its kinetic energy is lower. A deficit of kinetic energy is interpreted as a potential energy of the cluster. Thus, velocity, kinetic and potential energies are interpreted as parameters of an internal  state of automaton, which can be changed when at least two automata interacts during a collision. 

There are two fundamental constants of the theory. Let $d$ be a size of particle (Imp), and $\tau$ be an interaction time.
All particles are indistinguishable, therefore have equal mass $m=1$. Mass of a cluster is equal to the number of particles in this cluster.  
Particles and clusters have uniform motion until a collision/emission. Collisions are absolutely inelastic and form cluster which life-time $t_c\geq \tau$.  During a collision/emission momentum conversation holds. After a collision, the rest of kinetic energy is frozen inside the cluster in a form of potential energy, while during emission this potential energy can be released. From the momentum conservation 
\begin{equation}\label{moment-cons-cluster-col}
n_1V_1+n_2V_2=(n_1+n_2)V_c
\end{equation}
it follows that a collision of two clusters 
results in the following increase of potential energy
\begin{equation}\label{epot-cons-cluster-col}
\Delta E_{pot}=\frac{n_1n_2(V_1-V_2)^2}{n_1+n_2}\,.
\end{equation}
However, there are a lot of possibilities for emission processes which are restricted by energy and momentum conservation. For instance, the potential energy can be redistributed inside a cluster by some explicitly specified rule (or even randomly). We start with the simplest situation, where asymptotically interactions act as a permutation in the velocity space. Thus asymptotic velocities coincide with the case of instantaneous elastic collisions.

\begin{figure}[H]
\centering
\subfloat[Subfigure 1 list of figures text][2-particle clusters]{
\includegraphics[width=0.4\textwidth]{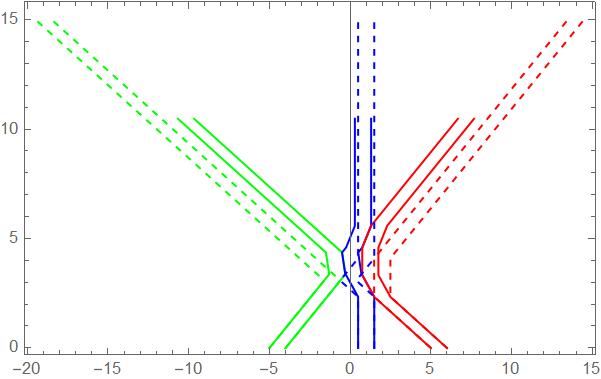}
\label{fig:3pts-frevsmic}}
\qquad
\subfloat[Subfigure 2 list of figures text][3-particle cluster]{
\includegraphics[width=0.4\textwidth]{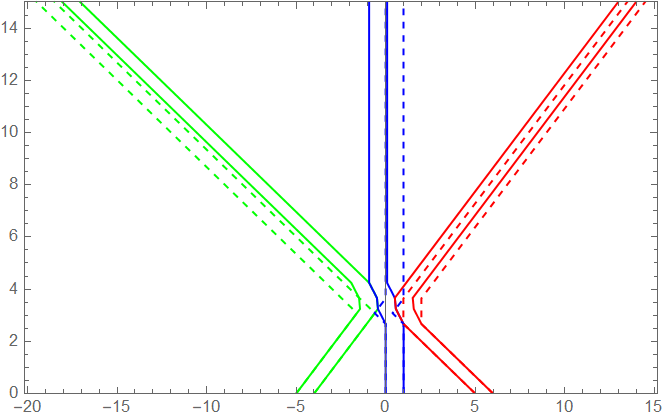}
\label{fig:3pts-frevsmic-cls}}
\caption{Three particle collisions of Imps compared with elastic collisions. Solid lines of the same color show edges of world-strips of Imps. Dashed lines correspond to instantaneous collisions.}
\label{fig:3pts}
\end{figure}

In figure \ref{fig:3pts} an example of 3-Imp collisions is compared with 3-particle absolutely elastic collisions.  

\section{Computer modeling of a multi-Imp one dimensional system}
One dimensional case has significant simplifications in a computational model. First of all, one can enumerate particles and by construction, the order of Imps stays the same during all evolution. The state of a cluster $C_{n_c}$ is completely characterized by
its left position $X_L$, its velocity $V_c$, number of particles $n_c$, $n_c$-tuples of initial velocities $\{v_1,v_2,\ldots v_{n_c}\}$, such that 
$$
V_c=\frac{1}{n_c}\sum\limits_{j=1}^{n_c}v_j\,.
$$ 
Finally, $n_c$-tuple of cluster formation temporal history $\{\tau_1,\tau_2,\ldots \tau_{n_c}\}$, $0\leq \tau_j\leq \tau$ defines when and how cluster explosion (or evaporation) could happen. 

Cluster evolution before a new collision or decay is just a uniform motion 
\begin{eqnarray}\label{def:cluster-propagation}
X_l(t_2)=X_l(t_1)+V_c(t_2-t_1)\,,\\
\{\tau_1(t_2),\tau_2(t_1),\ldots \,, \tau_{n_c}(t_2)\}=\{\tau_1(t_1)-(t_2-t_1),\tau_2(t_2),\ldots \tau_{n_c}(t_2)-(t_2-t_1)\}\,.
\end{eqnarray}
Note that only temporal variables $\tau_1$, $\tau_{n_c}$ at boundaries propagate.    

Let $\{v_1,v_2,\ldots v_{n_c}\}$ be velocities of Imps before the cluster formation. The energy of the system is conserved
\begin{equation}
E=\sum\limits_{j=1}^{n_c}v_j^2=n_cV_c^2+\sum\limits_{j=1}^{n_c}u_j    
\end{equation}
and the role of potential energy is to store information about the kinetic energies before cluster formations. 
We define the $n_c$-tuple of potential energies  $\{u_1,u_2,\ldots u_{n_c}\}$ as follows 
$$
u_j=v_j^2-V_c^2\,.
$$
This is not a unique option, however it is the simplest one. This definition implies that each Imp stores information about its initial kinetic energy. 
When cluster splits to separate Imps, potential energy is transformed to the kinetic energy of the Imps. Note also that we define the following interaction rule: let two interacting Imps store values $v_1,v_2$ before the interaction, when interaction is completed they exchange these values 
$$\{v_1\},\{v_2\}\to \{v_1,v_2\} \to \{v_2\},\{v_1\}$$
which resembles properties of elastic collisions.

Let us define rules of combination of clusters $$\{\ldots, C_{n_1},C_{n_2},\ldots\}\to \{\ldots\,, C_{n_1+n_2},\ldots\}$$ 
and emission of Imps by a cluster 
$$\{\ldots \,, C_{n_c},\ldots\}\to \{\ldots\,, C_1,C_{n_c-1}\,,\ldots\}$$
By dots we denote other clusters which are not involved into reconfiguration and freely propagate. 

To define which process (absorption or emission) will happen we compare time until emission $t_{e}=\text{min}(\tau_1,\tau_{n_c})$ and time until next collision $t_c=\text{min}\left(\frac{G_L}{V_L},\frac{G_R}{V_R}\right)$. Here $G_{L,R}$ are gaps which separate cluster from neighbors, and $V_{L,R}$ are relative velocities of neighbors and cluster ($L,R$ indicates neighbor from the left/right). 
\subsection{Emission}
Let $t_e=\tau_1<t_c$. We first apply free cluster propagation \eqref{def:cluster-propagation} from current time to $t\to t+t_e$. Than we separate our cluster to a free Imp and a cluster with $n_c-1$ Imps
\begin{equation}
\{u_1,u_2,\ldots u_{n_c}\}\to \{u_2,\ldots u_{n_c}\} 
\end{equation}

\subsection{Absorbtion}
Note that energy and momentum conservation allow us to determine without an ambiguity only two particle processes. If three or more particles collide and form a cluster, distribution of potential energy between particles of a cluster is ambiguous. In the simplest case, each two-particle collision leads to the exchange of potential energies after the interaction time $\tau$. As a result, emitted particles after complete cluster decay  will have the same kinetic energies as particles before cluster formation. In the limit $\tau \to 0$ corresponding model will reproduce classical billiard with absolutely elastic collisions between particles.

\section{Cluster formation and propagation}
\begin{figure*}
\centering
\includegraphics[scale=0.4]{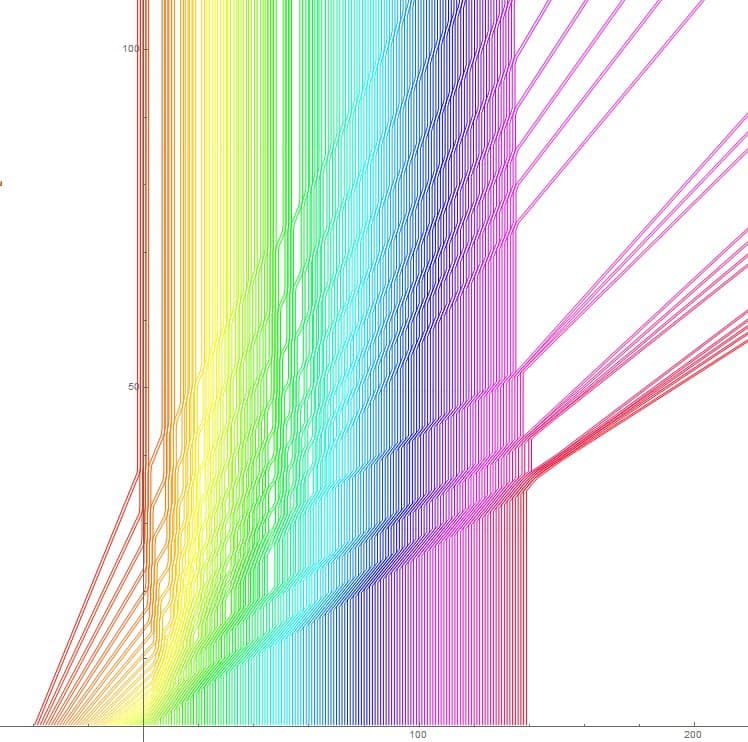}
\caption{A cluster formation and propagation through the medium.}\label{fig:cluster-formation}
\end{figure*}

Consider following initial conditions for a system of $N_m$ Imps inside a large box of size $L$. 
We choose the coordinate system where the gas of Imps is in the rest. Suppose that there exist a single Imp with large velocity, $v>0$. Let $l$ be the average free path in the gas, $l=L/N_m$. Large velocity of a Imp implies that the average time until a new collision with free Imps $t=l/v<2\tau$ is smaller than doubled interaction time. Therefore, each collision will join a new Imp to the initial one thus forming a growing cluster. Initial velocity is positive, therefore new Imps are joined from the right boundary of the cluster. At the same time, one Imp should be emitted from the left boundary at each time step $\tau$. If initial velocity $v$ is large, number of new collisions from the right boundary will be grater than the number of emitted Imps leading to the cluster growth. By momentum conservation, the velocity of the cluster will decrease with its growth. A growth of the cluster continues until its velocity drop up to {\it the critical velocity} $c=l/\tau$.

Let $n_c$ is a current number of Imps in the cluster. Since the gas is in the rest, the average momentum of free Imps equals to zero. Therefore the cluster velocity 
$$
v_c=\frac{v}{n_c}>c
$$
defines the number of collisions during single time step
$$
RC=\frac{v_c\tau}{l}\,.
$$
As a result, formation (growth) of cluster can be estimated 
by the following equation 
$$
\delta n_c=RC-1=\frac{v}{c}\frac{1}{n_c}-1
$$
which can be written in a discrete form
$$
n_c[T+1]=n_c[T]+\frac{v}{c}\frac{1}{n_c[T]}-1\,,\qquad n_c[0]=1\,.
$$
We also can consider its continuous asymptotic 
$$
n_c'[t]=\frac{v}{c}\frac{1}{n_c[t]}-1\,,\qquad n_c[0]=1\,.
$$
From the discrete equation it follows that during the first time step $\tau$ the cluster grows to its maximal (average) size from 1 to $[v/c]$ Imps. Afterwards its average velocity $\langle v_c\rangle=c$. Note that velocity fluctuations are possible due to fluctuations of number of captured and emitted particles. 
We can estimate these fluctuations as follows
$$
\delta v_c \propto \frac{c}{n_c+1}=\frac{c^2}{v+c}\propto \frac{c^2}{v}
$$
\begin{equation}\label{def:velocity-fluctuations}
\frac{\delta v_c}{v_c}\propto\frac{c}{v}=\frac{1}{n_c}
\end{equation}
Thus relative fluctuations of cluster velocity are inversely proportional to the size of cluster.

It should be noted, that due to the growth and decay of clusters in the medium of Imps, statistical properties of such a gas could be different compared to an ideal gas. In general, clusters of different sizes are present. In figure 
we present results of simulations of a system with 100 Imps inside a box. Horizontal line represents initial kinetic energy of Imps, and highly oscillated line represent potential energy (thus, it is a measure of how many Imps are combined into clusters).
\begin{figure*}
\centering
\includegraphics[scale=0.4]{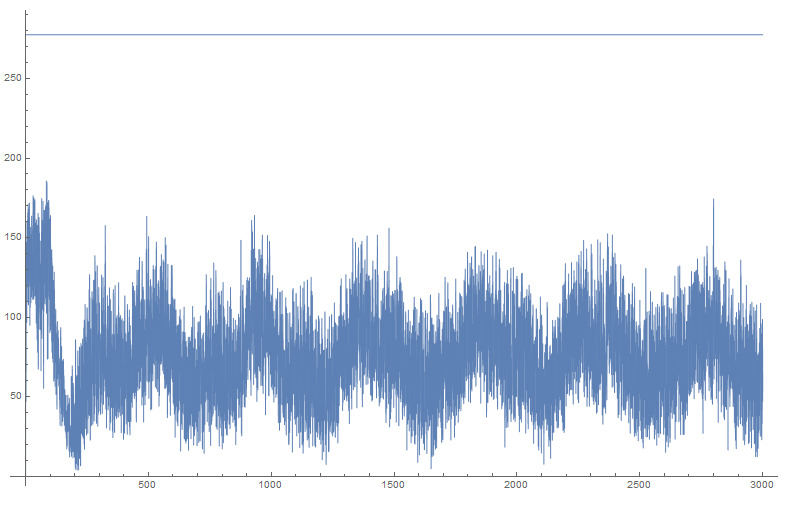}
\caption{Potential energy oscillations when a gas of Imps is confined to a box.}\label{fig:potential-energy}
\end{figure*}




\section{One dimensional Dirac equation}

In original Feynman and Hibbs formulation \cite{feynman2010quantum}:
"Suppose a particle moving in one dimension can go only forward or backward at the velocity of light." Let $\epsilon$ be a time between reversals of paths directions, and $R$ be a number of such reversals. Then, the amplitude of a path is defined as 
\begin{equation}\label{def:feynmanamplitude}
    \phi=(i \epsilon)^R
\end{equation}
and the weight $N(r)$ in the propagator  
\begin{equation}
K(x_1,x_2)=\sum_R N(r)(i \epsilon)^R
\end{equation}
counts all possible paths from $x_1$ to $x_2$ with $R$ reversals.
Moreover, the propagator is splitted into 4 elements $K_{\pm,\pm}(x_1,x_2)$ where symbols $\pm$   are related with the initial and final direction of velocity. The resulting matrix propagator coincides with the Green function of one-dimensional Dirac equation 
\begin{equation}
   \left( i\hbar \partial_t +i\hbar c \sigma_1\partial_x +mc^2 \sigma_ 0\right)K(x_1,x_2;t_1,t_2)=0\,,
\end{equation}
where
\begin{equation}
  \sigma_1=\begin{pmatrix}
0 & 1 \\
1 & 0
\end{pmatrix} \,,\quad  \sigma_0=\begin{pmatrix}
1 & 0 \\
0 & -1
\end{pmatrix}\,,\quad  K=\begin{pmatrix}
K_{++} & K_{+-} \\
K_{-+} & K_{--}
\end{pmatrix}
\end{equation}

Let us compare this relativistic stochastic walks with our model. We suppose that there exists single cluster inside a gaseous state of Imps.  
A gaseous medium of Imps stabilizes a cluster by instant collisions with its boundaries. In the equilibrium, the growth and evaporation of cluster on boundaries compensate each other. In the center of mass coordinate system the cluster stochastically interact with Imps from left and right boundaries. These interactions are responsible for a space-time picture of the motion, that is free Imps form a spatial lattice for stochastic cluster walks: average free path determines lattice spacing. 
\begin{figure*}
\centering
\includegraphics[scale=0.12]{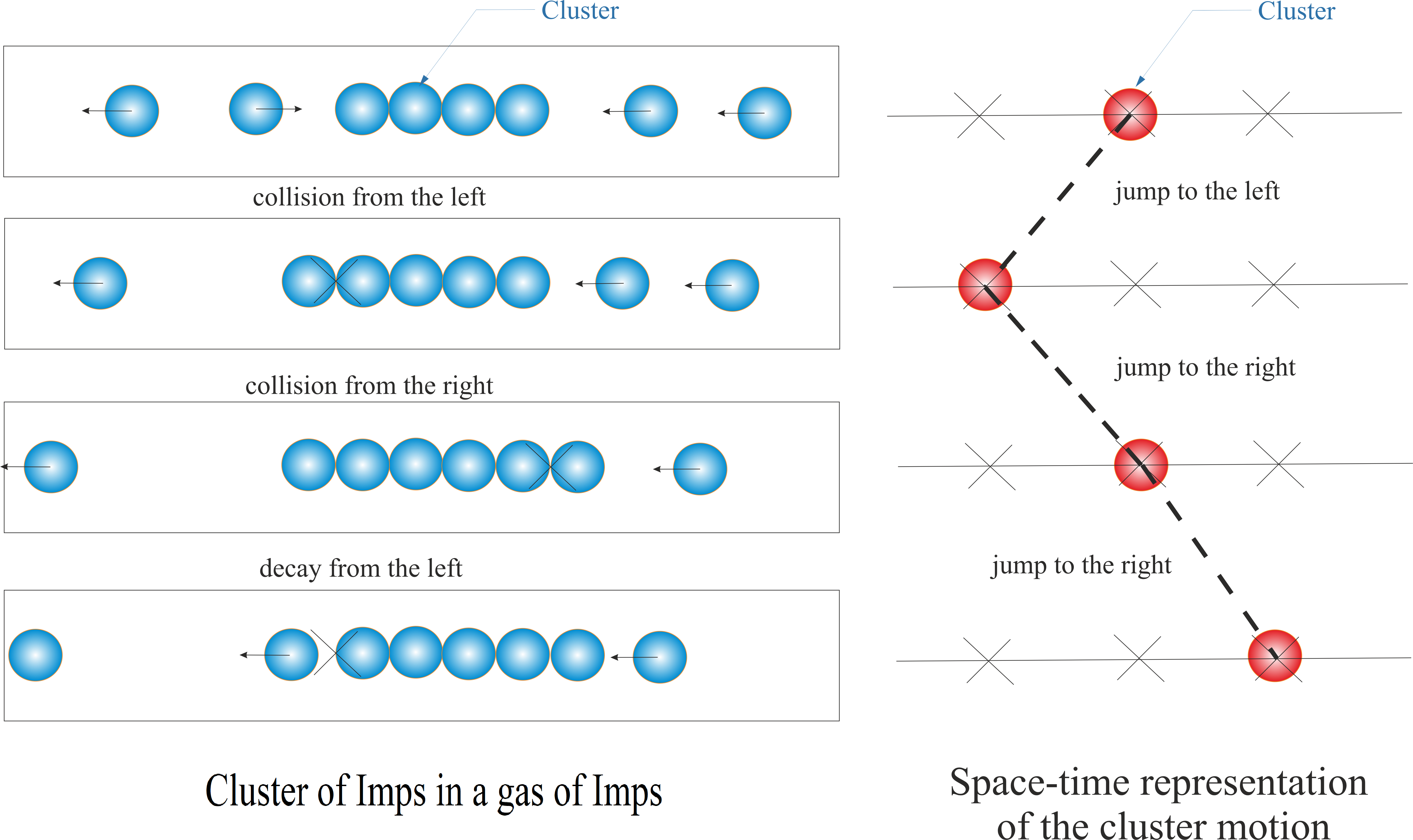}
\caption{Capture and decay as a relativistic walks on Feynman chessboard.}\label{fig:cap-decay-walks}
\end{figure*}
As a result, random walks of a cluster resembles a brownian motion. However,  a cluster has two boundaries and there are two types of interaction between a cluster and an environment (capture of a Imp and emission of a Imp). 
Therefore, a natural description of random walks using transitions between 4 states (capture from the left, capture from the right, emission from the left, emission from the right) appears.


 In the figure \ref{fig:cap-decay-walks} we establish a connection between the cluster Brownian motion in the gas of Imps and discrete jumps of a particle in the Feynman chessboard model. Originally, Feynman amplitude \eqref{def:feynmanamplitude}  was postulated to give desired propagator. Further analysis of this model revealed its connection with diffusion process and telegraph equation by analytical continuation. However, in \cite{ord1996schrodinger} it was shown that already classical random walks lead to both diffusion/Schr\"{o}dinger equations and telegraph/Dirac equations. In the latter work, quantum equations appeared as a classical equations which describe correlations in a multi-particle ensemble.
 the Schr\"{o}dinger and Dirac equations were derived as descriptions of
correlations induced on space-time geometry by the binomial distribution \cite{ord1996schrodinger}.

 Note, that \cite{ord1996schrodinger} introduces 4 internal states to a particle related with change of its velocity:
 
{\it We shall be working in 1+1 dimensions on a lattice with spacings $\delta$ and $\epsilon$ in
space and time respectively. At each lattice site a Brownian particle makes a binary
choice of whether to go left or right at the next step.
To keep track of the correlations in the trajectories we will assign states to the
particles as they move between lattice sites. States one and three will correspond to
particles moving to the right while states two and four will correspond to left
moving particles. States one and three will be distinguished by the parity of the
path. That is, states one and three will be separated by an odd number of transitions
from right moving to left moving (see Fig. 1). Similarly states two and four
will be separated by an odd number of left-to-right transitions.
}

 In the case of Brownian motion of a cluster such states can be associated with a capture/decay on the left/right edges the cluster. Thus, following \cite{ord1996schrodinger} and taking into account the existence of the speed limit (stability of the cluster implies that its velocity is equal to mean free path divided by $\tau$) we get both telegraph and Dirac equations to describe random walks of a cluster in a gas of Imps, where telegraph equation describes probabilities and Dirac equation describes correlations. 

Key ingredient of \cite{ord1996schrodinger} in the derivation of Dirac equation is the existence of a characteristic time scale. In the Kac
model each particle persists at constant speed $c$ along it's current direction for an average time $\langle t \rangle$ of say $1/\mu$. In our model this time-scale appears from the cluster velocity fluctuations \eqref{def:velocity-fluctuations} considered  in the cluster reference frame. We get $\langle t \rangle=1/n_c$, which defines an asymmetry of the binary choice of direction. If $p$ and $q$ correspond to persistence and
change in direction of cluster motion respectively in the binomial configuration generator then the
expected time of the first collision 
is
\begin{equation}
    \langle t \rangle =\sum_{k=1}^\infty k p^kq \tau
\end{equation}
By comparison with $\langle t \rangle= 1/n_c+\text{o}(\tau) $ to lowest order in $\tau$ we then must have $p=1-n_c \tau$ and $q=n_c\tau$, thus arriving to Eq. (29) of \cite{ord1996schrodinger} and finally to the Dirac equation.   
 
In the MCA model 1+1 dimensional lattice appear dynamically. Its spatial nodes correspond to the positions of free Imps and its temporal 
nodes correspond to collisions and emission of Imps and clusters. When dynamic of a cluster is considered, it is natural to consider 
only capture and emission of Imps by the cluster to form temporal nodes. Four internal states which track trajectories correlation 
are represented by the following processes: left capture, right emission, left emission, right capture.

Note that a process started with an emission of a Imp from the left and terminated by a  capture of a Imp from the right corresponds to the $K_{++}$ matrix element of the propagator.

%

\section{Conclusions}
In the present work, a mathematical model of a identical MCA with a resonance interaction on the line has been introduced. Such a model can be considered as an intermediate step between an ideal gas and a real gas. Already simple interaction which just temporarily combine independent particles in a cluster gives rise to various interesting structures. For example, in one dimensional gas of Imps confined in a box one can find clusters with small and large number of particles. Detailed studies of underlying statistical physics are beyond the content of our initial research.

We further studied a situation, where single cluster moves in a gas of Imps.
 The cluster is stable when its growth and evaporation compensate each other. This condition connects the density of gas and frequency of collisions between Imps. 
 When space and time are considered as emergent from the mutual interactions of Imps, displacement of a cluster to the closest Imp occurs during time $\tau$, which defines a critical speed). As a result, a stable state of a cluster looks like a particle moving in one dimension that can only go randomly forward or backward at the critical speed.
In \cite{kac1974stochastic}, such a stochastic model were used to derive telegraphers equation. Latter, \cite{ord1996schrodinger} extended results of \cite{kac1974stochastic} and deduced Dirac equation as equation which describes correlations between trajectory twists of a Brownian motion in a medium with a critical (maximal) speed. 

The same mathematics can be applied to our model in the considered regime. Thus a motion of a cluster which is formed from cellular automaton in a uniform gas of the same cellular automaton leads to the one-dimensional Dirac equation in the limit $\langle l\rangle\to 0$, $\langle t\rangle\to 0$, $\langle l\rangle/\langle t\rangle\to c$. Note that the critical speed (velocity of light) in the resulting equation is defined by the stability condition of the cluster.

The interpretation of the resulting Dirac equation on the original work \cite{ord1996schrodinger} is based on a classical multi-particle dynamics. However, in a fashion of Imp model, the Dirac equation describes a continuous limit  for the dynamic of a single cluster in a multi-particle environment, where single particle configurations are unobservable. Thus, a standard single particle probabilistic interpretation can be restored by tracing out gas of Imps which provide space-time lattice (and, at the same time, random generator).  

Here, probability comes from observable environment, and unknown history of cluster formation. It is thus unknown which side and when will emit an Imp. Thus even fixing an initial position of a cluster, there are different sequences of capture/emission processes which lead to a probability distribution for possible relative position of a cluster with respect to environment. This looks like a wave-packet spreading restricted by critical speed. 

On the other hand, imagine that we want to take a discrete Feynman model to approximate the propagator of the Dirac equation using the Monte Carlo method. Then, we need to run a lot of simulations tracking random walks on the lattice. This can be done using the gas of Imps surrounding the wandering particle - it turns out that it will act as a generator of random walks. And the rules for the growth and decay of clusters will lead to the presence of a maximum propagation velocity, so that the dynamics of Brownian motion will obey not the heat equation, but the telegraph equation.

\section{Acknowledgments}
These studies were supported by Mark Kopherstein.

\end{document}